\documentclass{article}

\usepackage{arxiv}

\usepackage[utf8]{inputenc} 
\usepackage[T1]{fontenc}    
\usepackage{hyperref}       
\usepackage{url}            
\usepackage{booktabs}       
\usepackage{amsfonts}       
\usepackage{nicefrac}       
\usepackage{microtype}      
\usepackage{lipsum}
\usepackage{amsmath,amssymb,amsfonts}
\usepackage{color}
\usepackage{algorithmic}
\usepackage{graphicx}
\usepackage{textcomp}
\usepackage{comment,color}
\usepackage{enumitem}
\usepackage{epsfig}
\usepackage{caption}
\usepackage{verbatim,bm,multicol,blindtext}
\usepackage{epstopdf,array,mathtools,url,subfigure}
\usepackage{multirow}

\title{Is it time to swish? Comparing activation functions in solving the Helmholtz equation using physics-informed neural networks}

\author{
  Ali Al-Safwan\\
  Department of Geosciences\\
  King Fahd University of Petroleum and Minerals
   \And
 Chao Song \\
  Department of Earth Science and Engineering\\
  Imperial College London
    \And
  Umair bin Waheed\\
  Department of Geosciences\\
  King Fahd University of Petroleum and Minerals
}

\begin{document}
\maketitle

\begin{abstract}
Solving the wave equation numerically constitutes the majority of the computational cost for applications like seismic imaging and full waveform inversion. An alternative approach is to solve the frequency domain Helmholtz equation, since it offers a reduction in dimensionality as it can be solved per frequency. However, computational challenges with the classical Helmholtz solvers such as the need to invert a large stiffness matrix can make these approaches computationally infeasible for large 3D models or for modeling high frequencies. Moreover, these methods do not have a mechanism to transfer information gained from solving one problem to the next. This becomes a bottleneck for applications like full waveform inversion where repeated modeling is necessary. Therefore, recently a new approach based on the emerging paradigm of physics informed neural networks (PINNs) has been proposed to solve the Helmholtz equation. The method has shown promise in addressing several challenging associated with the conventional algorithms, including flexibility to model additional physics and the use of transfer learning to speed up computations. However, the approach still needs further developments to be fully practicable. Foremost amongst the challenges is the slow convergence speed and reduced accuracy, especially in presence of sharp heterogeneities in the velocity model. Therefore, with an eye on exploring how improved convergence can be obtained for the PINN Helmholtz solvers, we study different activation functions routinely used in the PINN literature, in addition to the swish activation function – a variant of ReLU that has shown improved performance on a number of data science problems. Through a comparative study, we find that swish yields superior performance compared to the other activation functions routinely used in the PINN literature.
\end{abstract}


\section{Introduction}

Forward modeling lies at the heart of most inversion algorithms. Solving the wave equation numerically constitutes the majority of the computational cost for applications like seismic imaging and full waveform inversion (FWI). Frequency-domain solutions of the wave equation allow a reduction in dimensionality as they provide solutions of the Helmholtz equation per frequency. This approach to model wavefields gained popularity with the rise of waveform inversion~\cite{pratt1999seismic}; however, such solutions require inverting a stiffness matrix that can become computationally intractable for large 3D models or for modeling high frequencies. Moreover, conventional algorithms do not have a mechanism to transfer information gained from solving one problem to the next. For example, the same amount of computational effort is needed to model wavefields corresponding to a small perturbation in the velocity model. This results in a computational bottleneck when repeated modeling is needed for updated velocity models, as in the case of FWI.

Recently,~\cite{alkhalifah2020wavefield} demonstrated the use of physics-informed neural networks (PINNs) to address some of the challenges associated with the conventional Helmholtz solvers, including the flexibility of modeling additional physics. By considering a homogeneous background model, for which the wavefield solution is obtained analytically,~\cite{alkhalifah2020wavefield} trained a neural network to solve for the scattered wavefield. Contrary to conventional deep learning approaches, instead of a labeled set, the training of the neural network is done through a loss function, which embeds the residual of the underlying Helmholtz equation. For an input given by a location in the computational domain, the neural network learns to map it to the scattered wavefield solution at the input location, thanks to the universal approximation theorem. The key enabler of PINNs is automatic differentiation that allows for fast and accurate derivative computation of the neural network's output w.r.t. to the inputs, which is needed to evaluate the loss function. By using machine learning techniques like transfer learning and surrogate modeling, the method can be used to speed up repeated wavefield computations, as suggested by~\cite{waheed2020eikonal} for the eikonal equation. The PINN-based Helmholtz solver has recently been extended for the vertically transversely isotropic (VTI) case~\cite{song2021solving}.

However, several challenges remain to be addressed in making the PINN-based Helmholtz solver fully practicable. Chief among them is the slow convergence of the neural network in the presence of sharp heterogeneities in the velocity model. Although fully-connected neural networks can approximate arbitrary functions, they have been shown to prioritize learning low frequency features of the underlying function, and hence exhibit a bias towards smooth functions. This poses difficulty in training them to approximate, for example, solutions corresponding to high frequencies or solutions for sharply varying velocity models that generate high-frequency features in the wavefield solution. These observations have been confirmed both in theory and practice~\cite{rahaman2019spectral,cao2019towards}. 

Therefore, in this study, we explore the role of activation functions in improving convergence speed for the PINN-based Helmholtz solver. We do a comparative study of different activation functions that are routinely used in the PINN literature, namely hyperbolic tangent (tanh), inverse tangent (atan), and exponential linear unit (elu). In addition, we consider another variant of the ReLU activation function, named `swish', which has shown superior performance over other activation functions on a number of challenging image classification and machine translation problems~\cite{ramachandran2017searching}. Through tests on a synthetic model, we find that the swish activation function shows markedly improved convergence rate and accuracy in solving the Helmholtz equation.

\section{Theory}

The Helmholtz equation in an acoustic, isotropic medium with constant density is given as:
\begin{equation}
    \left(\nabla^2 + \frac{\omega^2}{v^2} \right) \, u(\mathbf{x}) = f(\mathbf{x}),
       \label{eq:HH}
\end{equation}
where $v$ is the wave velocity, $\omega$ is the angular frequency.
~The goal is to solve equation~\eqref{eq:HH} for the complex valued wavefield $u(\mathbf{x}) = \{u_r,u_i\}$ as a function of the angular frequency $\omega$. Then the time domain solution can be found using inverse Fourier transform of the superposition of the mono-frequency solutions.
To avoid the problem of sampling near the source location,~\cite{alkhalifah2020wavefield} suggests solving the Lippmann-Schwinger form of the Helmholtz equation: 
\begin{equation}
    \left(\nabla^2 + \frac{\omega^2}{v^2} \right) \, \delta u = -\omega^2 \, \delta m \, u_0,
       \label{eq:LS}
\end{equation}
for the scattered wavefield $\delta u = u - u_0$, where $u_0$ is the background wavefield satisfying equation~\eqref{eq:HH} for the background velocity $v_0$, and $\delta m = \frac{1}{v^2}-\frac{1}{v_0^2}$ is the velocity model perturbation. By using a background model with a constant velocity in which we can calculate the background wavefield $u_0$ analytically, we use equation~\eqref{eq:LS} to calculate the scattered wavefield $\delta u$ per frequency, from which we find the solution to the Helmholtz equation \eqref{eq:HH} as:
\begin{equation*}
    \begin{aligned}
    u_r & = \delta u_r + u_{0r},\\
    u_i & = \delta u_i + u_{0i}.\\
    \end{aligned}
\end{equation*}

To solve equation~\eqref{eq:LS} using a deep neural network (DNN), we consider a network with two neurons in the input layer for the spatial coordinates ($x$,$z$), two output neurons for the real and imaginary parts of the scattered wavefield, and a number of hidden layers. The second derivatives of the outputs ($\delta u_r, \delta u_i$) w.r.t. the inputs ($x$,$z$) can be computed using automatic differentiation. We seek to minimize the following loss function:
\begin{equation}
    \mathfrak{L} = \frac{1}{N}\sum_{i=1}^{N} || \omega^2 m^{(i)}  \delta u_r^{(i)}  + \nabla ^2  \delta u_r^{(i)}  + \omega ^2 \delta m   ~u_{0r}^{(i)}  ||_2^2  +  || \omega^2 m^{(i)}  \delta u_i^{(i)}  + \nabla ^2  \delta u_i^{(i)}  + \omega ^2 \delta m   ~u_{0i}^{(i)}  ||_2^2,
    \label{eq:loss}
\end{equation}
for the real and imaginary parts of the scattered wavefield $(\delta u_r, \delta u_i)$. 
Here $N$ is the number of training samples, which consists of a randomly selected collocation points from the computational domain, and $i$ is the sample index. The role of this loss function is to impose the validity of the Helmholtz equation on a given number of training points. The set of network parameters $\boldsymbol{\theta}^*$ that minimize the loss function given in equation \eqref{eq:loss} on this set of training samples can be found by solving the following optimization problem:
\begin{equation}
\boldsymbol{\theta ^*} = \arg\min_{\boldsymbol{\theta}} \mathfrak{L}(x^{*},z^{*}; \boldsymbol{\theta}).
\label{eq:optm}
\end{equation}
Once the DNN is trained, we evaluate it on a regular grid to obtain the predicted scattered wavefields, which are then added to the background wavefields to obtain the final wavefield solution.

\section{Activation functions}

The choice of activation function in DNNs has a significant impact on the training dynamics and convergence of the network. Activation functions are sources of non-linearity as they transform the neural network function to a superposition of non-linear basis functions. Choosing the right basis function can make a significant impact in efficiently representing a function. Therefore, in this study, we consider activation functions that are routinely used in the PINN literature (atan, tanh, and elu) in addition to the swish activation function, which is given as $f(x) = x \cdotp sigmoid(x)$. Swish is a smooth, non-monotonic function that has consistently matched or outperformed ReLU on DNNs applied to a variety of challenging problems, including image classification and machine translation. It is believed that the reason for this improvement is that swish is able to better alleviate the problem of vanishing gradients during backpropagation.

\section{Results}

\begin{figure}[ht!]
\begin{center}
\includegraphics[width=0.34\textwidth]{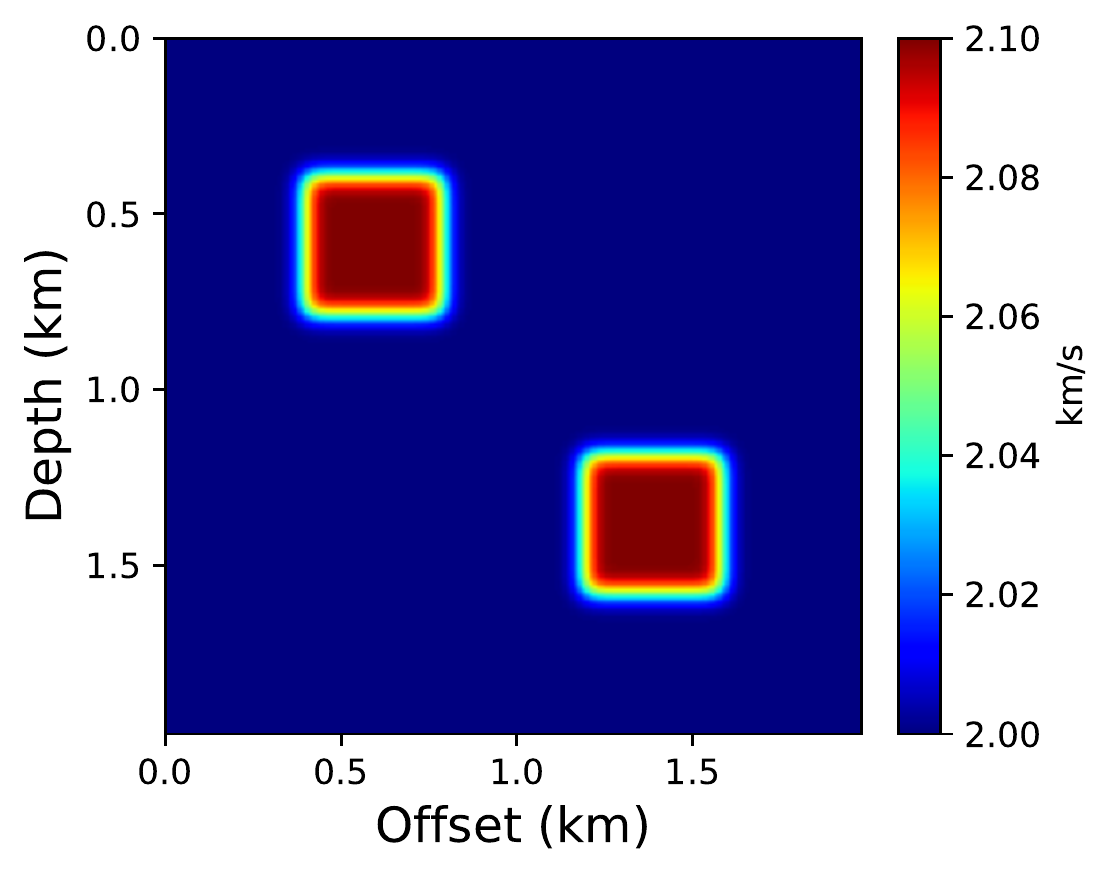}
\end{center}
\caption{
A two-scatter velocity model taken from~\cite{alkhalifah2020wavefield} for the tests. The background has a homogeneous velocity of 2~km/s.}%
\label{fig:model}
\end{figure}

In this section, we present a comparative study of different activation functions to solve equation~\eqref{eq:LS} for a two-scatter velocity model, shown in Figure~\ref{fig:model}. We keep all the hyper-parameters fixed to ensure fairness of the comparison. We use a DNN architecture with 8 hidden layers and 20 neurons in each layer, as used by~\cite{alkhalifah2020wavefield} for the considered velocity model. We randomly select 5000 locations from the computational domain and train the PINN model using the Adam optimizer for 15,000 epochs with mini-batch optimization having a batch size of 256. The PINN model is implemented using the SciANN package~\cite[]{haghighat2021sciann} -- a high-level Tensorflow wrapper for scientific computations.

\begin{figure}[ht!]
\begin{center}
\includegraphics[width=0.6\textwidth]{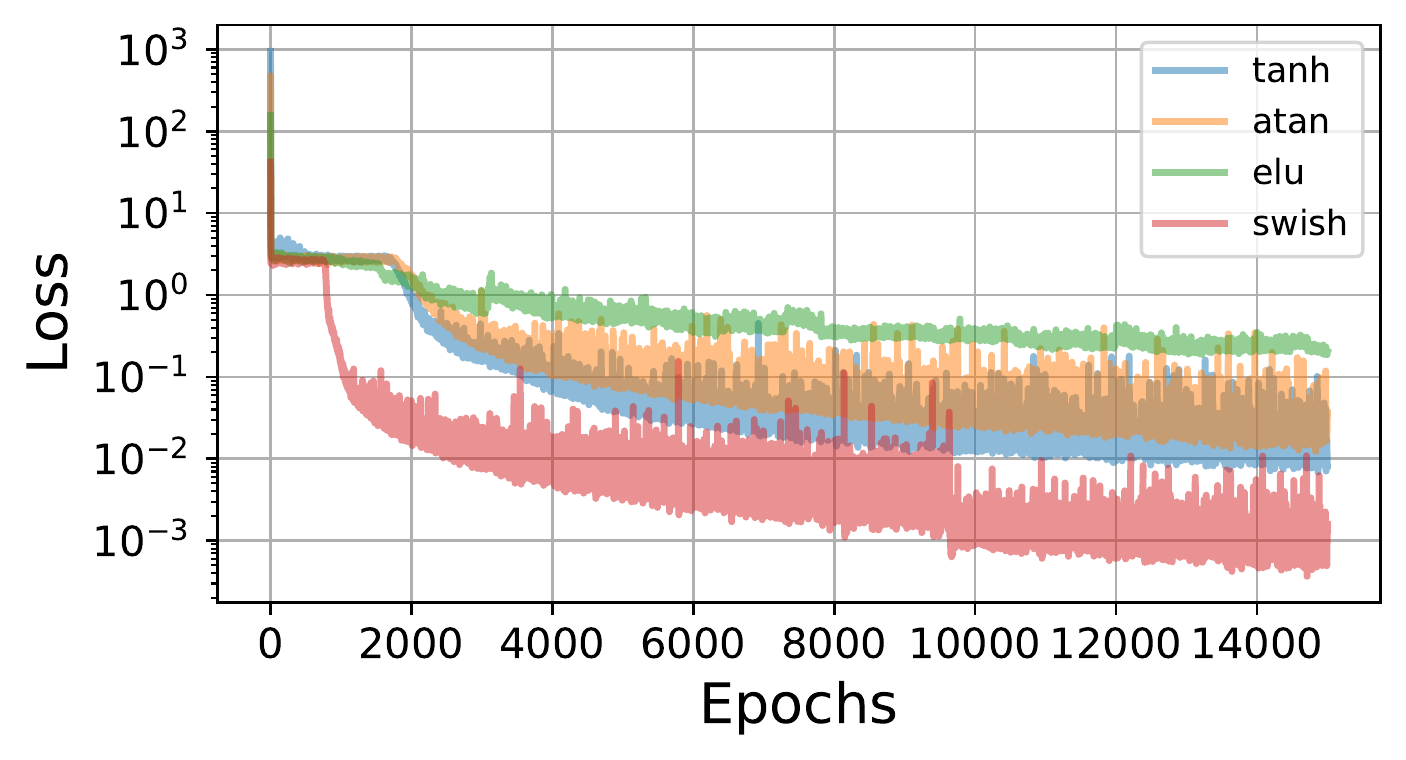}
\end{center}
\caption{
Loss history for training of the PINN model using different activation functions in the hidden layer.}%
\label{fig:loss}
\end{figure}

Figure~\ref{fig:loss} shows the loss history for training the PINN model using different activation functions considered. We observe that elu performs the worst with tanh and atan yielding similar performances, while swish has a markedly improved convergence compared to all the considered activation functions. For all activation functions, the convergence plateaus after an initial reduction of the loss value; however, swish recovers faster than the other activation functions. This can be observed by comparing the loss curves for the first 2000 epochs. We also tabulate the $L_2$ and $L_\infty$ errors for both the real and imaginary components of the wavefield solution using these activation functions (see Table~\ref{table1}). We clearly observe improved accuracy for the swish activation function compared to the rest.

\begin{figure}
  \centering
  \subfigure[]{\includegraphics[width=0.24\textwidth]{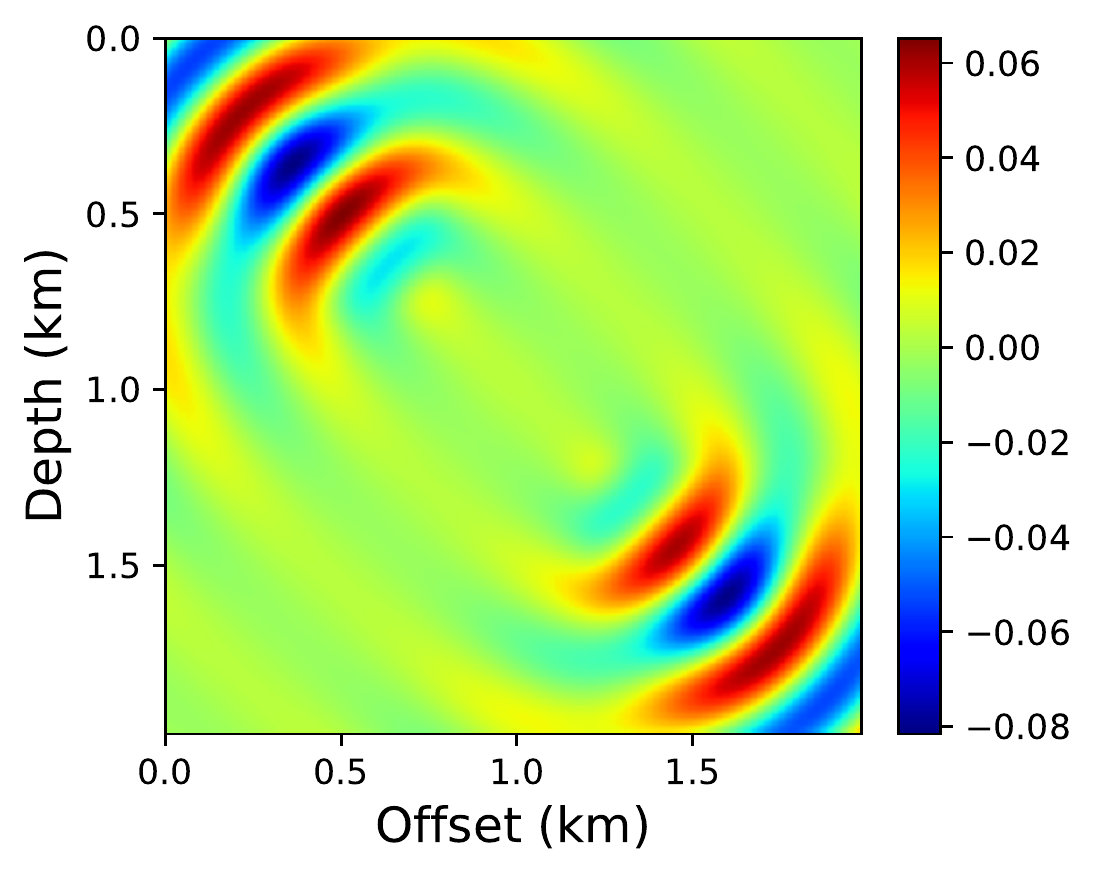}}
  \subfigure[]{\includegraphics[width=0.24\textwidth]{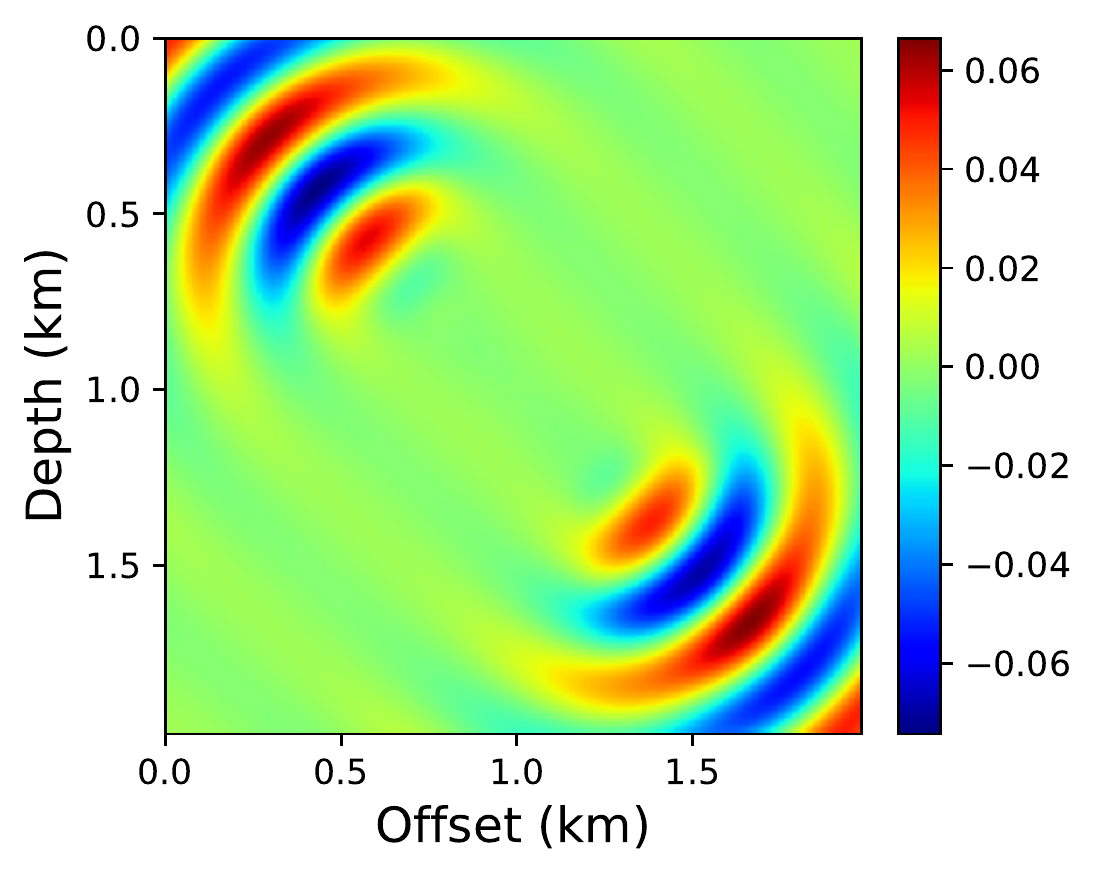}}
  \subfigure[]{\includegraphics[width=0.24\textwidth]{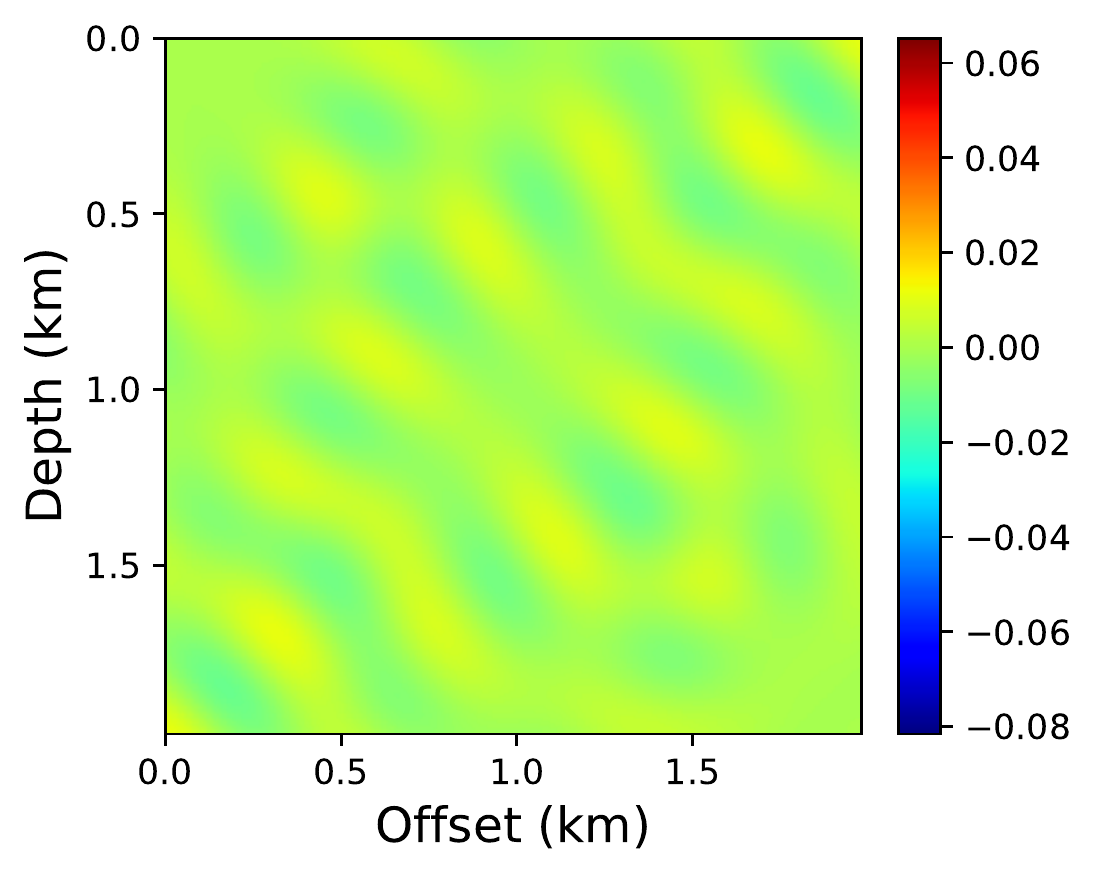}}
  \subfigure[]{\includegraphics[width=0.24\textwidth]{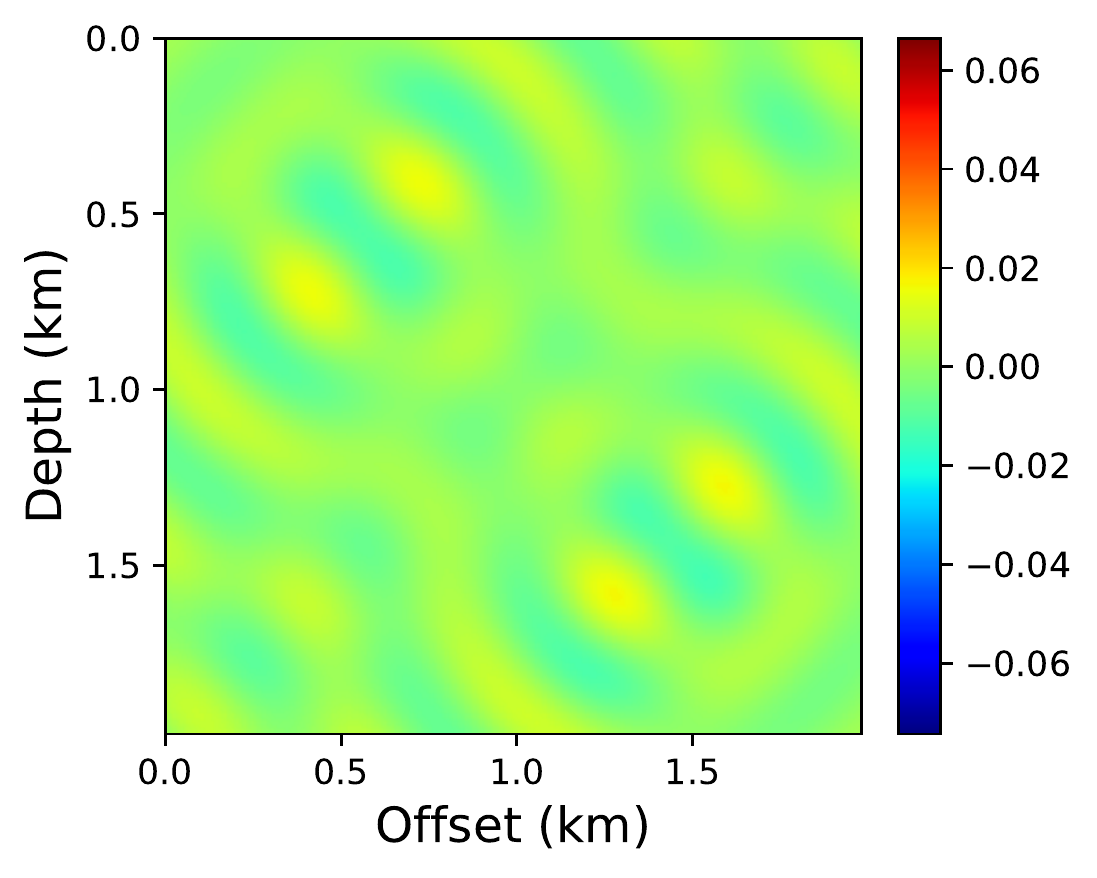}}
  \caption{The predicted scattered wavefield solution by the PINN model for the real (a) and imaginary (b) parts. Also shown are the wavefield errors for the real (c) and imaginary (d) parts.}
  \label{fig:solutions}
\end{figure}

\begin{table}
\centering
\begin{tabular}{|c|c|c|c|c|} \hline
\multirow{2}{*}{\begin{tabular}{c}Activation  functions\end{tabular}} & \multicolumn{2}{c|}{$L_2$} & \multicolumn{2}{c|}{$L_\infty$}\\ \cline{2-5}
 & Real & Imag. & Real & Imag. \\ \hline
tanh & $4.01\times 10^{-5}$ & $5.51\times 10^{-5}$ & $1.31\times 10^{-2}$ & $1.90\times 10^{-2}$ \\ \hline
atan & $4.15\times 10^{-5}$ & $5.53\times 10^{-5}$ & $1.09\times 10^{-2}$ & $1.81\times 10^{-2}$ \\ \hline
elu & $5.20\times 10^{-5}$ & $6.17\times 10^{-5}$ & $1.26\times 10^{-2}$ & $1.86\times 10^{-2}$ \\ \hline
swish & $3.65\times 10^{-5}$ & $4.7\times 10^{-5}$ & $9.86\times 10^{-3}$ & $1.58\times 10^{-2}$ \\ \hline
\end{tabular}
\caption{$L_2$ and $L_\infty$ errors for the real and imaginary parts of the PINN Helmholtz solution obtained using different activation functions.}
\label{table1}
\end{table}

In Figure~\ref{fig:solutions}, we show the scattered wavefield solution obtained from the PINN model trained using the swish activation function. We also plot the wavefield errors on the same color scale. The overall errors are considerably smaller than for the solutions computed using other activation functions as confirmed by Table~\ref{table1}.

 \section{Conclusions}

 We studied the routinely used activation functions from the PINN literature in solving the Helmholtz equation. In addition, we also considered the swish activation function, which is a variant of ReLU. Swish had shown improved performance compared to other activation functions in training DNNs for a variety of data science problems, thanks to its robustness against the problem of vanishing gradients. We also observed superior convergence using the swish activation function in solving the Helmholtz equation. Although these results are preliminary, they demonstrate the potential of selecting better activation functions to improve the convergence of PINN solvers. To focus on the activation functions in this study, we only used the Adam optimizer for all cases. However, to converge faster, typically the optimization process begins with a first-order optimizer like Adam and then switches to a second-order optimizer like L-BFGS after a suitably low value of the loss function. Seeing the loss curves, it can be noted that swish allows early switching to a second-order optimizer compared to other studied activation functions, which would compound its efficacy.
 
 \section{Acknowledgments}
 \vspace{-0.15cm}
 We extend gratitude to Prof. Tariq Alkhalifah and Dr. Ehsan Haghighat for helpful discussions.

\bibliographystyle{unsrt}  
\bibliography{references}  


\end{document}